\documentclass[a4paper,12pt]{article}
\usepackage{amssymb,latexsym}

\makeatletter \@addtoreset{equation}{section} \makeatother

\begin{document}

\begin{titlepage}

\thispagestyle{empty}

\begin{flushright}
\hfill{CERN-PH-TH/2004-010} \\
\end{flushright}

\vspace{35pt}

\begin{center}{ \LARGE{\bf
c--Map,very Special Quaternionic Geometry and Dual Ka\"hler
Spaces.}}

\vspace{60pt}

{\bf  R. D'Auria $^\diamond$, S. Ferrara$^{\dag\flat}$, M.
Trigiante $^\diamond$}

\vspace{15pt}

$^\dag${\it CERN, Physics Department, Theory Division, CH 1211
Geneva 23,
Switzerland}\\
$^\flat$ {\it Istituto Nazionale di Fisica Nucleare (INFN),
Laboratori Nazionali di Frascati, Italy}\\[1mm] {E-mail:
sergio.ferrara@cern.ch}

\vspace{15pt}

$^\diamond${\it Dipartimento di Fisica, Politecnico di Torino \\
C.so Duca degli Abruzzi, 24, I-10129 Torino, and Istituto
Nazionale di Fisica Nucleare, Sezione di Torino,
Italy}\\[1mm] {E-mail: riccardo.dauria@polito.it}\\[1mm] {E-mail: mario.trigiante@polito.it}

\vspace{50pt}

{ABSTRACT}

\end{center}

\medskip

We show that for all very special quaternionic manifolds a
different  $N=1$ reduction exists, defining a K\"ahler Geometry
which is ``dual'' to the original very special K\"ahler geometry
with metric $G_{a\bar{b}}= - \partial_a \partial_b \ln V$
($V=\frac{1}{6}d_{abc}\lambda^a \lambda^b \lambda^c$). The dual
metric $g^{ab}=V^{-2} (G^{-1})^{ab}$ is K\"ahler and it also
defines a flat potential as the original metric. Such geometries
and some of their extensions find applications in Type IIB
compactifications on Calabi--Yau orientifolds.

\end{titlepage}

\newpage

\baselineskip 6 mm

\section{Isometries of dual quaternionic manifolds}

One of the basic constructions in dealing with the low energy
effective Lagrangians of Type IIA and Type IIB superstrings is the
so called $c$--map \cite{cfg}, which associates to any Special
K\"ahler manifold of complex dimension $n$ a ``dual'' quaternionic
manifold
of quaternionic dimension $n_H=n+1$.\\
In particular it was shown \cite{fs} that ``dual'' quaternionic
manifolds always have at least $2n+4$ isometries: one scale
isometry $\epsilon_0$ and $2n+3$ shift isometries
$\beta_I,\alpha^I,\epsilon_+ $ ($I=0, \cdots, n $), whose
generators close a Heisenberg algebra \cite{adfft}:
\begin{equation}\label{commu}
 [\beta^I, \epsilon^+]= [\alpha_I, \epsilon^+]=0;
  \quad\,\, [\beta^I,\alpha_J]=\delta^I_J\epsilon^+;\quad\,\,
 [\epsilon^0,\alpha_I]=\frac {1}{2}\alpha_I;\quad\,\, [\epsilon^0,\beta^I]=
 \frac{1}{2}\beta^I;\quad\,\, [\epsilon^0,\epsilon^+] = \epsilon^+
\end{equation}
The corresponding generators can be written according to their
$\epsilon^0$ weight as \cite{alek,cecotti,dwvp1,dwvp2}:
\begin{equation}\label{weight}
\mathcal{V}=\mathcal{V}_0+\mathcal{V}_{\frac
{1}{2}}+\mathcal{V}_1.
\end{equation}
 However it was shown in \cite{dwvp1,dwvp2} that when the Special
 K\"ahler manifold has some isometries, then some ``hidden symmetries'' are generated
 in the $c$--map spaces
 which are classified by $\mathcal{V}_{-1}, \mathcal{V}_{-\frac
 {1}{2}}$, with
 \begin{equation}\label{dim}
 dim(\mathcal{V}_{-1})\leq 1; \quad dim(\mathcal{V}_{-
\frac{1}{2}})\leq 2n+2.
 \end{equation}
  In particular, for a generic very special geometry, with a
 cubic polynomial prepotential
 \begin{equation}\label{prep}
 F(z)= \frac{1}{48}d_{abc}z^a z^b z^c
\end{equation}
with generic $d_{abc}$, with no additional isometries, it was
shown that:
\begin{equation}\label{dim2}
dim(\mathcal{V}_{-1})=0;\quad \quad dim(\mathcal{V}_{-
\frac{1}{2}})=1;\quad \quad dim(\mathcal{V}_0)=n+2.
\end{equation}

Since the isometries of a generic very Special Geometry of
dimension $n$ are $n+1$, the dual manifold has then $3n+6$
isometries, where the $n+2$ additional isometries lie, $n+1$ in
$\mathcal{V}_0$, denoted by $\omega_I, (I=0,\cdots,n)$, and one
$\hat{\beta}_0$ in $\mathcal{V}_{- \frac{1}{2}}$. For symmetric
spaces the upper bound in equation (\ref{dim}) is saturated so
that ${\dim } G_Q={\dim } G_{SK}+4n+7$ where $ G_{SK}$ and $ G_Q$
are the isometry groups of the Special K\"ahler and Quaternionic
spaces respectively.
\section{The very Special $\sigma$--model Lagrangian and its $N=1$
reduction}

The quaternionic ``dual'' $\sigma$--model for a generic Special
Geometry was derived in \cite{fs} by dimensional reduction of a
$N=2$ Special Geometry to three dimensions. By adapting the
conventions of \cite{fs} to those of \cite {dwvp1} and \cite
{bghl} we call the special coordinates $z^a$ as $z^a=x^a+iy^a$ and
define:\begin{eqnarray}\label{VK} V&=&\frac {1}{6}(\kappa
yyy)\equiv \frac {1}{6}\kappa \quad \quad (\kappa yyy)=d_{abc}y^ay^by^c\\
\nonumber \kappa_a &=&d_{abc}y^by^c ;\quad \quad
\kappa_{ab}=d_{abc}y^c
\end{eqnarray}
The $2n+4$ additional coordinates are denoted by $\zeta^I\equiv
(\zeta^0,\zeta^a)$, $\tilde{\zeta}_I \equiv
(\tilde{\zeta}_0,\tilde{\zeta_a}),D,\tilde{\Phi}$.\par The
$\alpha^I,\,\beta_I$ isometries act as shifts on the $2n+2$
coordinates $\zeta^I,\,\tilde{\zeta}_I$:
\begin{equation}
    \delta\zeta^I=\alpha^I\,\,;\,\,\,\,\,\delta\tilde{\zeta}_I=\beta_I
\end{equation}
while the $\omega^a$ shift isometries of the special geometry,
$\delta x^a=\omega^a$, act as duality rotations on the
$\zeta^I,\,\tilde{\zeta}_I$ symplectic vector:
\begin{equation}
    \delta\left(\matrix{\zeta \cr \tilde{\zeta}}\right)=\left(\matrix{A & 0\cr C &
    -A^T}\right)\,\left(\matrix{\zeta \cr \tilde{\zeta}}\right)
\end{equation}
with
\begin{equation}
    A=\left(\matrix{0 & 0\cr \omega^a &
    0}\right);\quad\quad C=\left(\matrix{0 & 0\cr 0 &
    3\,d_{abc}\,\omega^c}\right).
\end{equation}
On the other hand the $\hat{\beta}_0$ isometry rotates $\zeta^a$
into $x^a$ so that the $x^a,\,\tilde{\zeta}_a$ variables are
related by quaternionic isometries.
 It is
immediate to see that the full $\sigma$--model Lagrangian
\cite{fs,dwvp1,bghl} is invariant under the following parity
operation $\Omega$:
\begin{equation}\label{parity} y^a\rightarrow y^a;\quad \quad
\tilde{\zeta_a}\rightarrow \tilde{\zeta_a};\quad \quad
\zeta^0\rightarrow \zeta^0;\quad \quad D\rightarrow
D;\end{equation}
\begin{equation}
 x^a \rightarrow -x^a;\quad \quad \zeta^a\rightarrow -\zeta^a;\quad
\quad \tilde{\zeta}_0\rightarrow -\tilde{\zeta}_0;\quad \quad
\tilde{\Phi}\rightarrow -\tilde{\Phi}
\end{equation}
so that, restricting to the plus-parity sector is a consistent
truncation, giving rise to the following Lagrangian for $2n+2$
(real) variables:
\begin{eqnarray}\label{lag1}
(\sqrt -g)^{-1} \mathcal{L}&=& -(\partial_{\mu}D)^2 -\frac
{1}{4}G_{ab}\,\partial_{\mu}y^a \partial^{\mu}y^b-
\frac {1}{8}e^{2D}\,V\,(\partial_{\mu}\zeta^0)^2\\
\nonumber &&-2e^{2D}\, V^{-1}\,(G^{-1})^{ ab}\partial_\mu
\tilde{\zeta}_a
\partial^\mu \tilde{\zeta}_b
\end{eqnarray}
where $G_{ab}=-\partial_a\partial_b \log V$. By a change of
variables we can decouple the $(D,\,\zeta^0)$ fields from the rest
as follows: define two new variables $(\Phi,\,\lambda^a)$:
\begin{equation}
V(y)\,e^{2D}=e^{2\Phi}\,\,;\,\,\,\,\,y^a=\lambda^a\,e^{\frac{\Phi}{2}}
\end{equation}
Thus it follows that $V(\lambda) e^{2D}=e^{\frac{\Phi}{2}}$ and
the Lagrangian becomes:
\begin{eqnarray}\label{lag2}
(\sqrt -g)^{-1} \mathcal{L}&=& -\frac {1}{4}(\partial_{\mu}\Phi)^2
-\frac {1}{8}e^{2\Phi}(\partial_{\mu}\zeta^0)^2 -\frac
{1}{4}G_{ab}\partial_{\mu}\lambda^a \partial^{\mu}\lambda^b\\
\nonumber &-&\frac {1}{4}(\partial_\mu \log V(\lambda))^2-2
V(\lambda)^{-2}(G^{-1})^{ ab}\partial_\mu \tilde{\zeta}_a
\partial^\mu \tilde{\zeta}_b
\end{eqnarray}
the $(\Phi,\,\zeta^0)$ part defines a ${\rm SU}(1,1)/{\rm U}(1)$
$\sigma$--model.\par The coefficient of the two terms in the
$\partial_{\mu}\lambda^a
\partial^{\mu}\lambda^b$ part combine into $-\frac{3}{2}\left( \frac{\kappa_{ab}}{\kappa}-
3\frac{\kappa_{a}\kappa_{b}}{\kappa^2}\right)$. \\We now define a
new variable $t_a=\frac{1}{2}\kappa_{ab} \lambda^b$ such that
$d\lambda^b=(\kappa^{-1})^{ba}\, t_a$ we obtain that
\begin{equation}
g^{ab}= -6\,\left( \frac{\kappa_{cd}}{\kappa}-
3\frac{\kappa_{c}\kappa_{d}}{\kappa^2}\right)(\kappa^{-1})^{ac}(\kappa^{-1})^{bd}=-\frac{6}{\kappa^2}
\left[(\kappa^{-1})^{ab}\kappa -3\lambda^a \lambda^b
\right]=\frac{36}{\kappa^2} (G^{-1})^{ab}
\end{equation}
Therefore in the $(t_a,\,\tilde{\zeta}_a)$ variables we finally
get
\begin{eqnarray}
(\sqrt{-g})^{-1} \mathcal{L}&=& -\frac
{1}{4}(\partial_{\mu}\Phi)^2 -\frac
{1}{8}e^{2\Phi}(\partial_{\mu}\zeta^0)^2 -\frac
{1}{4}g^{ab}\partial_{\mu}t_a \partial^{\mu}t_b\\
\nonumber &&-2 g^{ab}\partial_\mu \tilde{\zeta}_a
\partial^\mu \tilde{\zeta}_b
\end{eqnarray}
Therefore by defining the complex variables
\begin{equation}
\eta_a=t_a+2\sqrt{2}i \tilde{\zeta}_a
\end{equation}
we get for the $2n$--dimensional $\sigma$--model:
\begin{eqnarray}
-\frac {1}{4}g(\Re \eta )^{ab}\left(\partial_\mu\Re \eta_a\,
\partial^\mu\Re \eta_b+\partial_\mu\Im \eta_a\,
\partial^\mu\Im \eta_b\right)&=&-\frac
{1}{4}g^{ab}\partial_\mu\eta_a\,
\partial^\mu \bar{\eta}_b
\end{eqnarray}
The previous Lagrangian is K\"ahler provided
\begin{equation}
g(t )^{ab}= \frac{\partial^2 \hat{K}}{\partial t_a\,\partial t_b}.
\end{equation}
This condition is achieved by setting $\hat{K}=-2\,\log
V(\lambda)$. Indeed
\begin{eqnarray}
\frac{\partial}{\partial t_a} \log
V&=&(\kappa^{-1})^{ac}\frac{\partial}{\partial \lambda^c} \log V=
3
\frac{\lambda^a}{\kappa}\nonumber\\
\frac{\partial^2 }{\partial t_a\,\partial t_b}\log V&=&
3\,\left[\frac{(\kappa^{-1})^{ab}}{\kappa} -3\frac{\lambda^a
\lambda^b}{\kappa^2} \right]=-\frac
{1}{2}\times\frac{36}{\kappa^2}\,(G^{-1})^{ab}
\end{eqnarray}
\section{Isometries of the $N=1$ reduction}
The $\sigma$--model isometries of the c--map, using the notations
of \cite{dwvp2} are parametrized by
\begin{equation}
\epsilon^+,\,\epsilon^0,\,\alpha^I,\,\beta_I,\,\omega^a,\,\omega^0,\,\hat{\beta}_0.
\end{equation}
The $N=1$ reduction projects out
$\epsilon^+,\,\alpha^a,\beta_0,\,\omega^a$, so the remaining
isometries are $n+4$, namely:
\begin{equation}
\beta_a,\,\omega^0,\,\epsilon^0,\,\alpha^0,\,\hat{\beta}_0.
\end{equation}
Three of the latter generate a ${\rm SL}(2,\mathbb{R})$ symmetry
(otherwise absent in generic dual quaternionic manifolds), the
others generate a shift symmetry in $\Im \eta_a$ and a scale
symmetry in the $\eta_a$ variables. The dual manifold has the same
isometries of the original Special K\"ahler. Since the
$\tilde{\zeta}_a$ variables are related to the $x^a$ variables by
quaternionic isometries, the two manifolds need in fact not be
distinct. Even though the $\tilde{\zeta}_a$ variables are related
to the $x^a$ variables by quaternionic isometries, the two
manifolds are in general distinct.  However, in the particular
case of homogeneous--symmetric spaces \cite{cvp}, it turns out
that the dual manifold coincide with the original one. The proof
of this statement will be given elsewhere.
\section{Connection with Calabi Yau orientifolds }
The c--map was originally studied in relation to the Type II A
$\rightarrow$ Type II B mirror map in Calabi--Yau
compactifications. In Calabi Yau orientifolds of Type II B strings
with D--branes present, the bulk Lagrangian is obtained combining
a world--sheet parity with a manifold parity which, for generic
spaces \cite{gktt}, is precisely doing the truncation we have
encountered in this note.\par For certain Calabi Yau manifolds
more generic orientifoldings are possible where the set of special
coordinates $z^A$ is separated in two parts with opposite parity,
$z_\pm^A$ ($n_++n_-=n$) such that \cite{ggjl}
\begin{eqnarray}
y_\pm &\rightarrow &\pm \,y_\pm\nonumber\\
x_\pm &\rightarrow &\mp \,x_\pm
\end{eqnarray}
and then consequently
\begin{eqnarray}
\zeta_\pm &\rightarrow &\mp \,\zeta_\pm\,\,\,;\,\,\,\,\,\,\zeta^0 \rightarrow \zeta_0\nonumber\\
\tilde{\zeta}_\pm &\rightarrow &\pm
\,\tilde{\zeta}_\pm\,\,\,;\,\,\,\,\,\,\tilde{\zeta}_0 \rightarrow
-\tilde{\zeta}_0
\end{eqnarray}
 However in this case one must demand
\begin{equation}
d_{++-}=d_{---}=0
\end{equation}
in order for the $N=1$ reduction to be consistent \cite{adf}.\par
In this case the $\sigma$--model Lagrangian acquires more terms
and can be symbolically written as:
\begin{eqnarray}\label{lagpm}
(\sqrt{-g})^{-1}\,{\cal L}&=&-(\partial D)^2-\frac{1}{4}\,G_{++}\,
(\partial y_+)-\frac{1}{4}\,G_{--}\, (\partial x_-)^2
-\frac{1}{8}\,e^{2D}\,V\,(\partial
\zeta^0)^2-\nonumber\\&&\frac{1}{8}\,e^{2D}\,V\,G_{--}\,(x_-
\partial\zeta^0-\partial\zeta_-)^2-\nonumber\\&&2
e^{2D}\,V^{-1}\,(G^{-1})^{++}\,(\partial
\tilde{\zeta}_++\frac{1}{8}\,d_{+--}\,x_- x_- \partial
\zeta^0-\frac{1}{4}\,d_{+--}\, x_- \partial \zeta_-)^2
\end{eqnarray}
where for the sake of simplicity space--time indices have been
suppressed from partial derivatives and contraction over them is
understood. In (\ref{lagpm}) $G_{++}$ is as before since
$d_{+++}\neq 0$,  $G_{+-}=0$ and
$G_{--}=-6\,(d_{--+}y_+)/(d_{+++}y_+y_+y_+)$.\par
 The total set of coordinates are:
 $y_+,\,x_-,\,\zeta_-,\,\tilde{\zeta}_+$ and ($\Phi,\,\zeta^0$).
 Since in this case some of the $y$ coordinates, namely $y_-$, have been
 replaced by $x_-$, the new variables define a K\"ahler manifold
 of complex dimension $n+1$ certainly distinct from the original
 one.
\par
There is an $N=4$ analogue of this dual $N=1$ geometries if we
consider different embeddings of $N=4$ supergravity into $N=8$.
This corresponds to Type II B on $T^6/\mathbb{Z}_2$ orientifold
with $D3$ or $D9$ branes (Type I string) or Heterotic string on
$T^6$. In all these cases the bulk sector corresponds to $[{\rm
SO}(6,6)/{\rm SO}(6)\times {\rm SO}(6)]\times [{\rm SU}(1,1)/{\rm
U}(1)]$ $\sigma$--model but the 15 axions in ${\rm SO}(6,6)/{\rm
SO}(6)\times {\rm SO}(6)$ are coming from $C_4,\,C_2,\,B_2$
\cite{fp,kst,dfv1,dfv2,dfgtv,aft1,aft2}.\par Also cases in which a
further splitting appears are realized if the orientifold
projection acts \cite{aft1} differently on $T^{p-3}\times T^{9-p}$
($p=3,5,7,9$). This is the analogue of the $y_\pm,\,x_\pm$
splitting \cite{ggjl}. In all these cases the  dual manifolds
coincide, as predicted by $N=4$ supergravity.
\section{Properties of the dual Special K\"ahler spaces and no--scale structure.}
The dual K\"ahler space, obtained by a $N=1$ truncation of the
(c--map) very special quaternionic space has a metric that
satisfies a ``duality'' relation with the original very special
K\"ahler space:
\begin{equation}
g_D^{ab}=\frac{1}{V^2}\, (G^{-1})^{ab}
\end{equation}
Moreover it can be shown that its affine connection is simply
related to the affine connection of original K\"ahler space:
\begin{equation}
\Gamma^D_d{}^{bc}=\frac{1}{V}\, (G^{-1})^{ca}\Gamma^b_{ad}.
\end{equation}
Actually in the one--dimensional case the two connections
coincide.\par These dual spaces are also no--scale
\cite{cfkn,bcf,cetal}. Indeed it is sufficient to prove that
\begin{equation}
\frac{\partial\hat{K}}{\partial\Re \eta_a}
(g^{-1})_{ab}\frac{\partial\hat{K}}{\partial\Re \eta_b}=3.
\end{equation}
But this is indeed the case since
\begin{equation}
\lambda^a G_{ab} \lambda^b =3.
\end{equation}
From a Type II B perspective, this was anticipated in \cite{tv}
\section{Concluding remarks.}
In this note we have shown that for an arbitrary very special
geometry, through the c--map, it is possible to construct a
``dual'' K\"ahler geometry which has a dual metric, it is K\"ahler
and it provides a dual no--scale potential. Recently such
constructions have found applications in Calabi Yau orientifolds
\cite{bbhl,ggjl} but the procedure considered here is intrinsic to
the four dimensional context.\par We have not shown that the final
Lagrangian is supersymmetric but, using the reduction techniques
of \cite{adf}, it can be shown that this is indeed the case. It is
reassuring that the ${\rm SL}(2,\mathbb{R})$ symmetry, related to
the Type II B interpretation, comes out in a pure four dimensional
context, thanks to the results of \cite{dwvp1,dwvp2}
\section{Acknowledgements}
We would like to thank M.A. Lled\'o and A. Van Proeyen for
enlightening discussions.\par
 The work of S.F. has been supported in
part by the D.O.E. grant DE-FG03-91ER40662, Task C, and in part by
the European Community's Human Potential Program under contract
HPRN-CT-2000-00131 Quantum Space-Time, in association with INFN
Frascati National Laboratories and R.D. and M. T. are associated
to Torino University.

\end{document}